\documentclass[%
 aip,
 amsmath,amssymb,
 reprint,%
]{revtex4-1}

\usepackage{graphicx}
\usepackage{dcolumn}
\usepackage{bm}

\usepackage[utf8]{inputenc}
\usepackage[T1]{fontenc}
\usepackage{mathptmx}
\usepackage{etoolbox}

\makeatletter
\def\@email#1#2{%
 \endgroup
 \patchcmd{\titleblock@produce}
  {\frontmatter@RRAPformat}
  {\frontmatter@RRAPformat{\produce@RRAP{*#1\href{mailto:#2}{#2}}}\frontmatter@RRAPformat}
  {}{}
}%
\makeatother
\begin{document}

\preprint{AIP/123-QED}

\title[Optimal design of fast topological pumping]{Optimal design of fast topological pumping}
\author{Xianggui Ding}
\affiliation{State Key Laboratory of Structural Analysis, Optimization and CAE Software for Industrial Equipment, Department of Engineering Mechanics, Dalian University of Technology, Dalian, 116023, China}

\author{Zongliang Du}%
\affiliation{State Key Laboratory of Structural Analysis, Optimization and CAE Software for Industrial Equipment, Department of Engineering Mechanics, Dalian University of Technology, Dalian, 116023, China}
\affiliation{Ningbo Institute of Dalian University of Technology, Ningbo, 315016, China}
\email{zldu@dlut.edu.cn}

\author{Jiachen Luo}
\affiliation{%
Department of Mechanical Engineering, Boston University, Boston, MA 02215, USA}

\author{Hui Chen}
\affiliation{Piezoelectric Device Laboratory, School of Mechanical Engineering and Mechanics, Ningbo University, Ningbo 315211, China}%
\email{chenhui2@nbu.edu.cn}

\author{Zhenqun Guan}
\affiliation{State Key Laboratory of Structural Analysis, Optimization and CAE Software for Industrial Equipment, Department of Engineering Mechanics, Dalian University of Technology, Dalian, 116023, China}
\affiliation{Ningbo Institute of Dalian University of Technology, Ningbo, 315016, China}

\author{Xu Guo}
\affiliation{State Key Laboratory of Structural Analysis, Optimization and CAE Software for Industrial Equipment, Department of Engineering Mechanics, Dalian University of Technology, Dalian, 116023, China}
\affiliation{Ningbo Institute of Dalian University of Technology, Ningbo, 315016, China}

\date{\today}

\begin{abstract}
Utilizing synthetic dimensions generated by spatial or temporal modulation, topological pumping enables the exploration of higher-dimensional topological phenomena through lower-dimensional physical systems. In this letter, we propose a rational design paradigm of fast topological pumping based on 1D and 2D time-modulated discrete elastic lattices for the first time. Firstly, the realization of topological pumping is ensured by introducing quantitative indicators to drive a transition of the edge or corner state in the lattice spectrum. Meanwhile, with the help of limiting speed for adiabaticity to calculate the modulation time, a mathematical formulation of designing topological pumping with the fastest modulation speed is presented. By applying the proposed design paradigm, topological edge-bulk-edge and corner-bulk-corner energy transport are successfully achieved, with 11.2 and 4.0 times of improvement in modulation speed compared to classical pumping systems in the literature. In addition, applying to 1D and 2D space-modulated systems, the optimized modulation schemes can reduce the number of stacks to 5.3\% and 26.8\% of the classical systems while ensuring highly concentrated energy transport. This design paradigm is expected to be extended to the rational design of fast topological pumping in other physical fields.
\end{abstract}

\maketitle

The investigation of topological insulators has stimulated broad interest among researchers in many disciplines, such as photonics,\cite{RN1,RN2,RN3,RN4} acoustics,\cite{RN5,RN6,RN7,RN8} and mechanics.\cite{RN9,RN10,RN11,RN12,RN13} The conventional strategy involves analogs to the quantum Hall effect,\cite{RN14,RN15,RN16} the quantum spin Hall effect,\cite{RN17,RN18,RN19} and the quantum valley Hall effect\cite{RN20,RN21,RN22} through breaking time-reversal symmetry and spatial symmetry to achieve robust energy transport with immunity to defects and backscattering. In fact, such topologically protected modes can also be pursued through synthetic dimensions, and previous researches have achieved higher-dimensional topological effects in lower-dimensional physical systems by exploiting synthetic dimensions in parameter space.\cite{RN23,RN24,RN25} In particular, it has been demonstrated in several physical systems that topological pumps can achieve edge-to-edge transitions of topological states through the dynamic evolution of parameters involved in space \cite{RN26,RN27,RN28,RN29,RN30,RN31} or time.\cite{RN32,RN33,RN34,RN35,RN36} In mechanical systems, topological pumping of edge states has been successfully realized in elastic discrete lattices and elastic plates via synthetic dimensions generated by spatial modulation.\cite{RN29,RN30} In addition, with synthetic dimensions generated by temporal modulation, it is also possible to construct elastic topological pumps in beams with stiffness modulated by piezoelectric patches or resonators and elastic lattices with stiffness modulated by rotary mechanisms.\cite{RN33,RN36,RN37} These researches highlight the potential of spatial or temporal pumps for robust control of elastic waves.

Since topological protection is often accompanied with adiabaticity requirements, a sufficiently long period  is required to suppress the loss of energy in the process of the fast-pumping evolution of edge states. Besides, the modulation parameters of current pumping systems are primarily determined by the experience of  designers,\cite{RN26,RN29,RN30,RN31,RN33,RN36} and there lacks of rational design process. Recently, topology optimization has been successfully applied to the rational design of topological insulators,\cite{RN38,RN39,RN40,RN41,RN42,RN43,RN44} showing many advantages over trial-and-error methods, such as obtaining highly concentrated topological edge mode and a larger operating bandwidth. In contrast to topological insulators, the coupling and modulation between unit cells are crucial for implementing the pumps. To the best of the authors' knowledge, there is no rational design work for pumping systems in the literature. In response to the above challenges, this letter develops a rational design paradigm for 1D and 2D pumping systems of elastic wave, achieving the fastest adiabatic evolution of edge and corner states by optimally designing the modulation parameters of the spring stiffness. Compared with the classical modulation parameters in the literature, the optimized modulation parameters can significantly improve the adiabatic modulation speed of the pumping systems.

\begin{figure}
\includegraphics[width=0.4\textwidth]{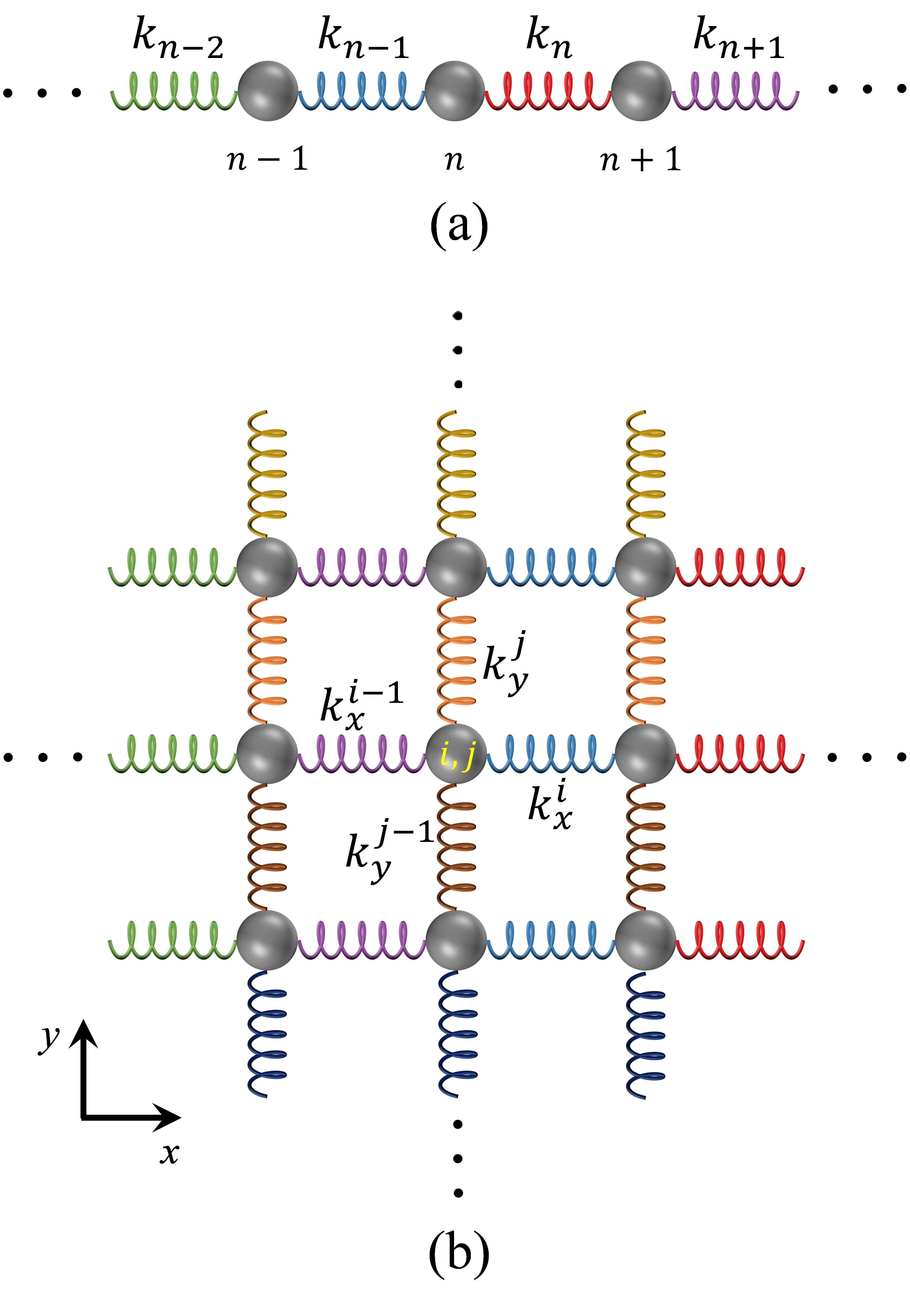}
\caption{\label{fig:1} Topological pumping systems. (a) 1D spring-mass lattice. (b) 2D square spring-mass lattice.}
\end{figure}

In this letter, we use 1D and 2D elastic lattices as illustrating examples to validate the effectiveness of the proposed design paradigm. As shown in Fig. \ref{fig:1}(a), the 1D system consists of 30 identical mass blocks of weight $m$, which are connected by equal-length springs with time-modulated stiffness. The stiffness of the $n$th spring is defined as:
\begin{equation}
k_n=k_0\left[1+\delta\cos{\left(2\pi nb+\phi\left(t\right)\right)}\right] %
\label{eq:1}
\end{equation}
Here, $k_0$ is a constant, and $\phi$ is the modulation phase, which satisfies $\phi(t)=\phi_0+\Omega t$ under linear modulation, where $\Omega$ is the modulation speed. $\delta$ and $b$ define the amplitude and frequency of stiffness modulation, and are set as part of design variables.

The square lattice of the specified 2D system contains 12 × 12 identical mass blocks of weight $m$, which are similarly connected by equal-length springs with time-modulated stiffness, as shown in Fig. \ref{fig:1}(b). The spring stiffness is specifically defined as: 
\begin{equation}
 k_{\alpha n}=k_0\left[1+\delta_\alpha\cos{\left(2\pi n b_\alpha+\phi_\alpha\left(t\right)\right)}\right] %
 \label{eq:2}
\end{equation}
where $\delta_\alpha$, $b_\alpha$ and $\phi_\alpha$ define the amplitude, frequency, and phase of the stiffness modulation in the $\alpha=x,y$ directions, respectively. Without loss of generality, we assume $\phi_x=\phi_y=\phi=\phi_0+\Omega t$, indicating that the phase values in the $x$-direction and $y$-direction are consistent over time.

\begin{figure*}
\includegraphics[width=0.9\textwidth]{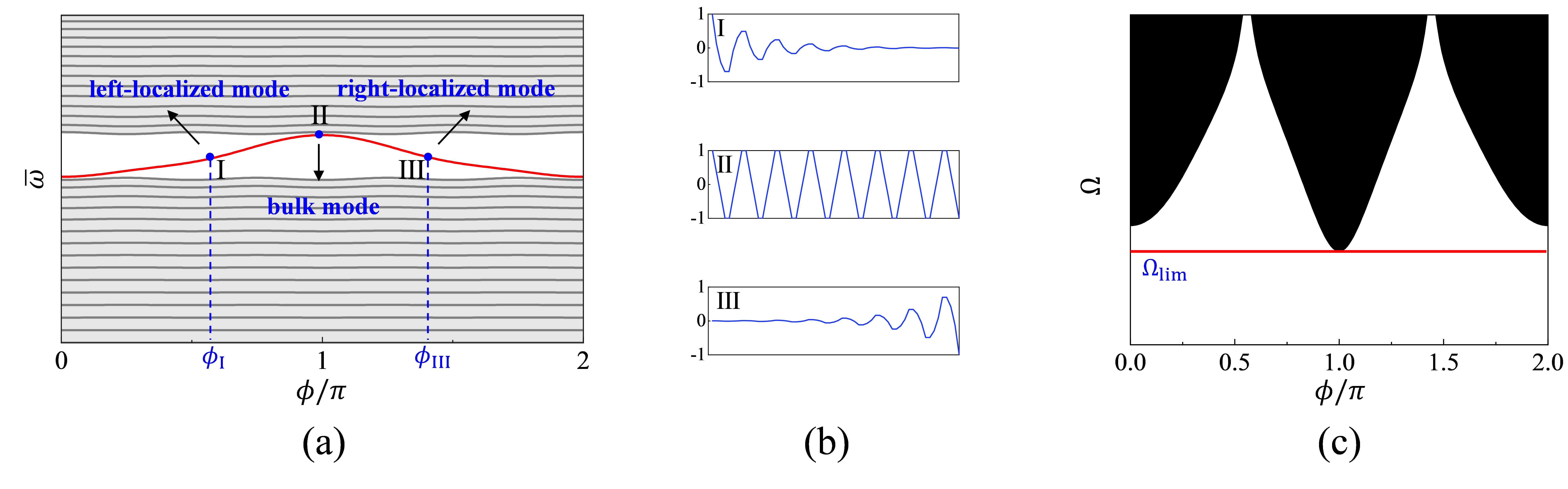}
\caption{\label{fig:2}  (a) Spectrum of a 1D elastic system, the shaded gray regions denote the bulk bands, and the red curve identifies the edge band. (b) Modes corresponding to the points marked in Fig. 2(a). (c) Graphical representation of the adiabatic (white region) and non-adiabatic (black region) transformations upon varying the modulation speed $\Omega$ and the phase $\phi$, while the horizontal red line corresponds to the limiting speed for adiabaticity.}
\end{figure*}

Since the spring stiffness varies with time, the motion of the system at a certain moment $t$ is determined by the corresponding instantaneous eigenvalue problem:
\begin{equation}
 \mathbf K\left(\phi\left(t\right)\right)\textbf{\textit{u}}=\omega^2\mathbf M\textbf{\textit{u}} %
 \label{eq:3}
\end{equation}
where $\mathbf{K}$, $\mathbf{M}$ denote the stiffness and mass matrices of the system, and $\omega$, $\textbf{\textit{u}}$ are the eigenfrequency and corresponding eigenmode, respectively. By defining a dimensionless frequency $\bar{\omega}=\sqrt{m\omega^2/k_0}$, Fig. \ref{fig:2}(a) shows the quasi-static spectrum of a 1D spring-mass system, i.e., $\bar{\omega}$ against the variation of $\phi$. The gray regions denote the bulk bands, and an edge band spanning the nontrivial gap is highlighted by the red line. Representative right-localized ($\rm I$), bulk ($\rm II$), and left-localized ($\rm III$) modes are displayed in Fig. \ref{fig:2}(b) to illustrate a transition of the edge state.

The design idea in this letter is to customize the edge band of the spectrum mentioned above (the red curve in Fig. \ref{fig:2}(a)) by optimization, i.e., to achieve the desired transformations of topological modes along the synthetic dimension by ensuring that the specified phases correspond to specified modes. As an illustration, the mathematical formulation for the 1D spring-mass lattice is as follows:
\begin{align}
\rm find\ \ \ \ &\mathit{d}=\left(b,\delta,\phi_{\rm I},\phi_{\rm III}\right)^\top \nonumber\\
\min\ \ \ \ &\mathcal{F}_{\rm r}(\phi_{\rm I})+\mathcal{F}_{\rm l}(\phi_{\rm III})+\mathcal{F}_{\rm b}(\pi)+T/\bar{T} \nonumber\\
\rm s.t.\ \ \ \ &\mathcal{F}_{\rm l}=H(V_{\rm l}-\mathrm{\Psi}_{\rm l}/\widetilde{\mathrm{\Psi}}) \nonumber\\
&\mathcal{F}_{\rm r}=H(V_{\rm r}-\mathrm{\Psi}_{\rm r}/\widetilde{\mathrm{\Psi}}) \nonumber\\
&\mathcal{F}_{\rm b}=H({V_{\rm b1}-\mathrm{\Psi}}_{\rm b}/\widetilde{\mathrm{\Psi}})+H(\mathrm{\Psi}_{\rm b}/\widetilde{\mathrm{\Psi}}-V_{\rm b2})
\label{eq:4}
\end{align}
where the first three terms of the objective function are designed to ensure that the modes belonging to the edge band correspond to the right-localized, left-localized, and bulk modes at phases $\phi_{\rm I}$,\ $\phi_{\rm III}$ and $\phi_ {\rm II}=\pi$, respectively. The fourth term is to achieve the shortest modulation time for topological pumping, where $\bar{T}$ is used for dimensionless treatment. $H(x)$ is the Heaviside function, $V_{\rm l}$, $V_{\rm r}$,  $V_{\rm b1}$, and $V_{\rm b2}$ are the threshold parameters for the concentration indicators set(see Appendix S1 for details).

In order to ensure the adiabatic properties of energy transport, we generalize Eq. (\ref{eq:3}) into a first order differential equation $\dot{\varphi} = \mathbf{H}\varphi$ with state vector $\varphi=[\dot{\textit{\textbf{u}}};\ddot{\textit{\textbf{u}}}]$. The Hamiltonian matrix $\mathbf{H}$ can be expressed as:
\begin{equation}
\mathbf{H}=\begin{bmatrix}
	 \mathbf{0} & -\mathbf{M}^{-1}\mathbf{K}(\phi(t))  \\
	\mathbf{I} & \mathbf{0}
	 \end{bmatrix}
  \label{eq:5}
\end{equation}
where $\mathbf{I}$ is the identity matrix. Then we can adopt the approximation conditions indicative of the adiabatic transition to achieve the limiting speed for adiabaticity:\cite{RN33,RN35,RN36,RN45,RN46}
\begin{equation}
\xi =\left| \frac {\left<L_p(t)\left|\frac{\partial \mathbf H}{\partial \phi}\right|R_q(t)\right>}{(\omega_p-\omega_q)^2}\Omega\right| \ll 1
\label{eq:6}
\end{equation}
where $p$, $q$ denote the state indexes involved in the modal transformation. In general, state $p$ is the topological edge state and state $q$ is the bulk state next to it since the amount of coupling between the two states decreases with the square of the distance $(\omega_p-\omega_q)^2$. $\left <L_p (t)\right|$ and $\left |R_q (t)\right>$ are the left and right eigenmodes corresponding to the eigenfrequencies $\omega_p$ and $\omega_q$, respectively. $\xi$ is a visual indication of the coupling between the two states mentioned above. Based on the condition Eq. (\ref{eq:6}), the limiting speed for adiabaticity corresponding to the modal transformation on the edge band in Fig. \ref{fig:2}(a) is shown in Fig. \ref{fig:2}(c). Obviously, the modulation speed should not exceed $\Omega_{\rm lim}$ involving linear modulation. Thus, the time mentioned in the mathematical formulation can be expressed as $T=(\phi_{\rm III}-\phi_{\rm I})/\Omega_{\rm lim}$.

\begin{figure*}
\includegraphics[width=0.9\textwidth]{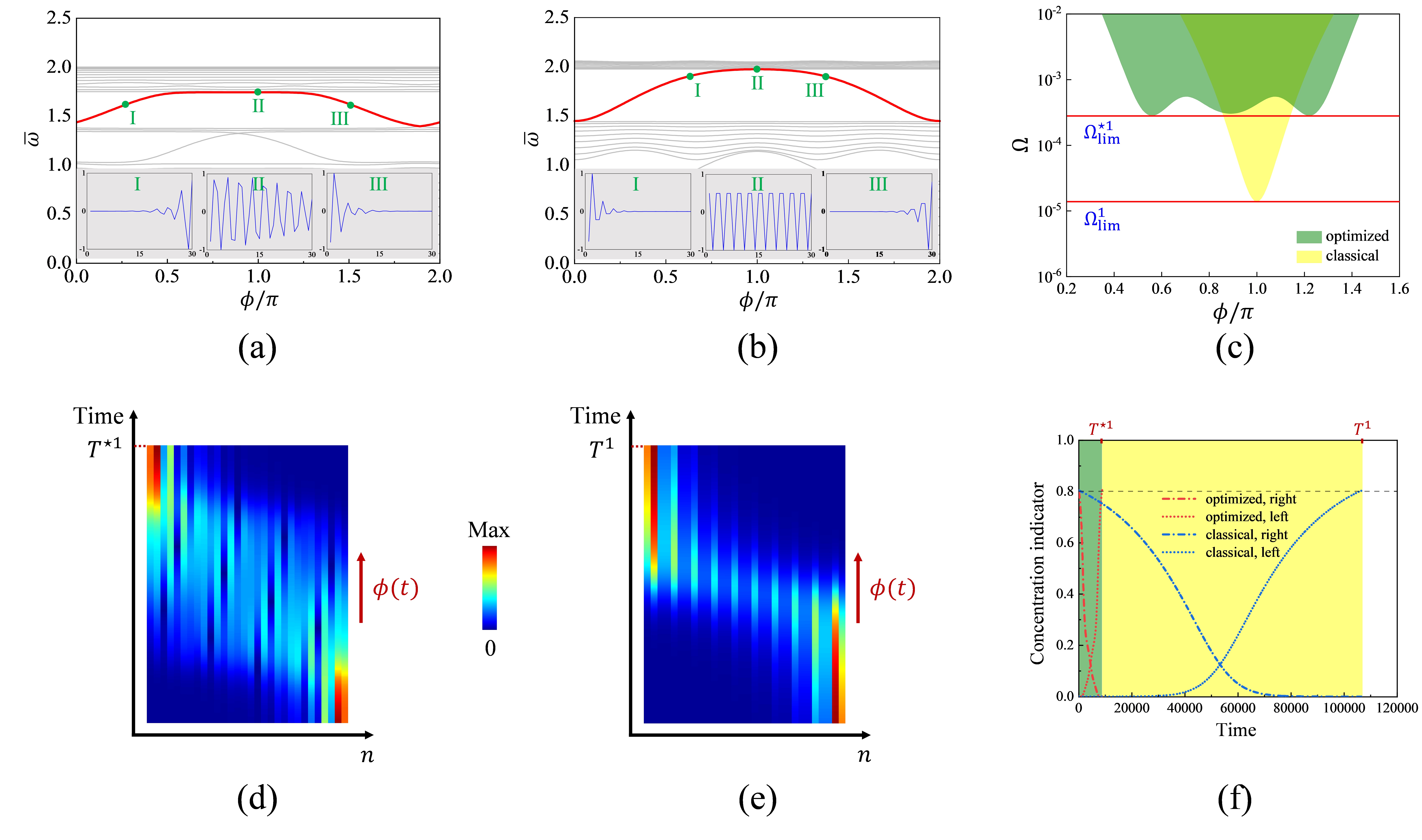}
\caption{\label{fig:3} (a) Spectrum of the 1D optimized system. (b) Spectrum of the classical system. (c) Adiabatic ($\xi\leq0.01$) and non-adiabatic ($\xi>0.01$) regions for the optimized (green) and classical (yellow) systems. Under narrow-band excitation (d) Displacement field of the optimized system. (e) Displacement field of the classical system. (f) Concentration indicators of edge modes over time in the optimized and the classical systems.}
\end{figure*}

By setting $\xi=0.01$, $b\in\left[0.01,1\right]$, and $\delta\in\left[0.01,1\right]$, the optimized modulation parameters $\mathit{d}_1^\star=\left(0.4370,0.4947,0.27\pi,1.51\pi\right)^\top$ are obtained by solving formulation (\ref{eq:4}) with the help of genetic algorithm.
Accordingly, spectrums of the 1D optimized pumping system $\mathit{d}_1^\star$ and classical pumping system\cite{RN30}($b=1/3,\delta=0.6$) are shown in Figs. \ref{fig:3}(a) and \ref{fig:3}(b), respectively. It can be found that both the optimized and classical systems can drive a transition of the edge state from right localized (I), to bulk (II), and then to left localized (III) by increasing phase from $\phi_{\rm I}$ to $\phi_{\rm III}$. Furthermore, the limiting speeds for adiabaticity and the modulation times can be obtained as $\Omega_{\lim}^{\star1}=1.4154\pi\times{10}^{-4}$, $T^{\star1}=8761$ and $\Omega_{\lim}^1=6.9498\pi\times{10}^{-6}$, $T^1=106478$, and the modulation time of the optimized system is only 1/12.2 of that of the classical system. Fig. \ref{fig:3}(c) displays adiabatic and non-adiabatic regions for the optimized and classical systems, and it can be found that the limiting speed for adiabaticity of the optimized system is significantly increased. To validate the results, right-localized edge modes are induced by applying a narrow-band excitation in the corresponding systems, and then the spring stiffness is modulated according to the optimized and classical modulation schemes, respectively. Figs. \ref{fig:3}(d) and \ref{fig:3}(e) show the displacement fields of the 1D optimized as well as the classical systems over time, where the color scale corresponds to the vibration amplitude. It can be seen that both systems achieve highly concentrated energy transport. In order to validate the adiabaticity of the pumps, Fig. \ref{fig:3}(f) shows the variations in the concentration indicators of edge modes (see Appendix S1 for details) in different systems. At the initial moment, the concentration indicator of the right-localized edge mode is around 0.8, which gradually decreases over time and finally tends to 0, while the concentration indicator of the left-localized edge mode gradually rises from 0 and finally reaches 0.8, indicating that there is no obvious energy leakage in either the optimized system or the classical system. However, the optimized system dramatically reduces the modulation time, which is highly consistent with the design objective.

\begin{figure*}
\includegraphics[width=0.9\textwidth]{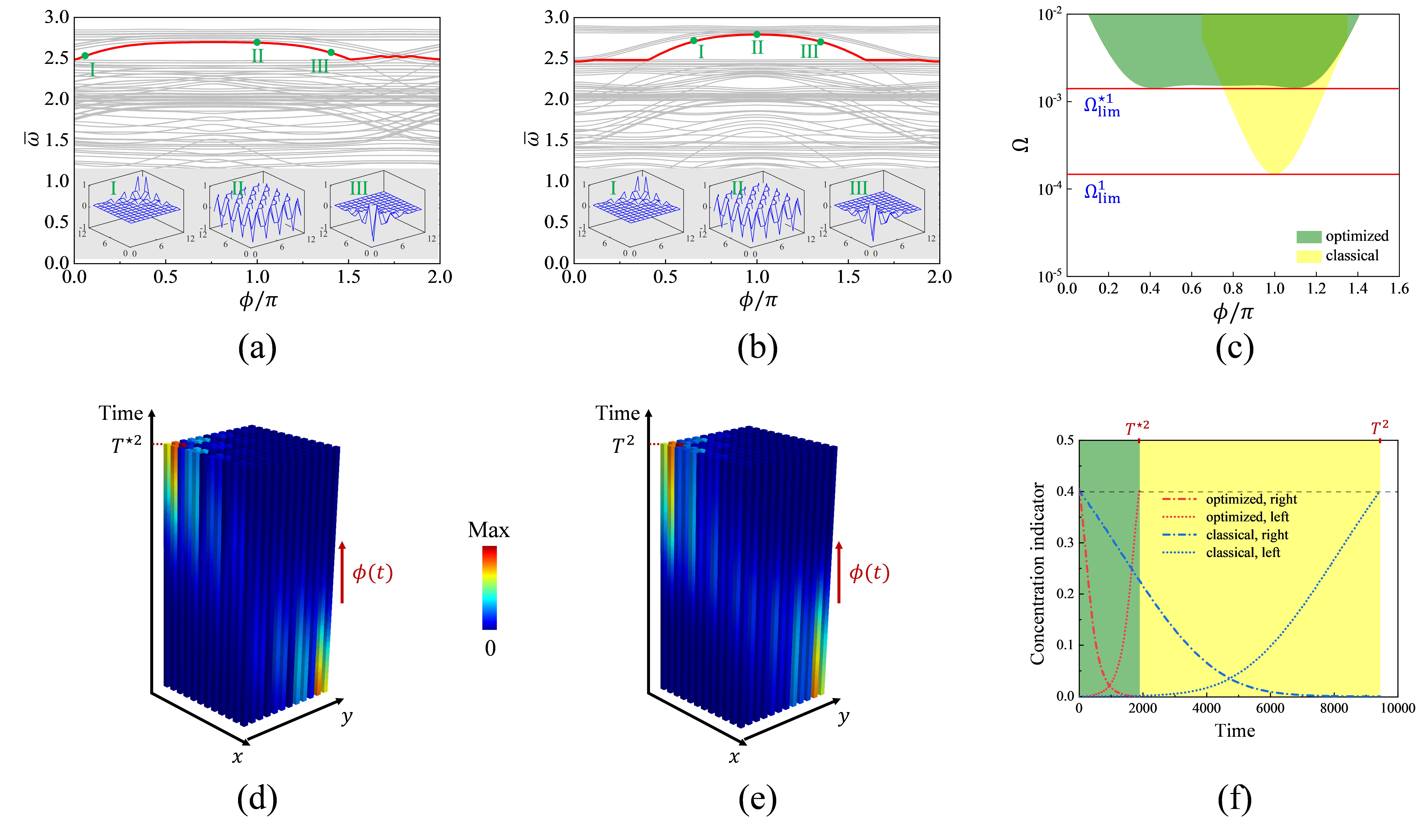}
\caption{\label{fig:4}  (a) Spectrum of the 2D optimized system. (b) Spectrum of the classical system. (c) Adiabatic ($\xi\leq0.01$) and non-adiabatic ($\xi>0.01$) regions for the optimized (green) and classical (yellow) systems. Under narrow-band excitation (d) Displacement field of the optimized system. (e) Displacement field of the classical system. (f) Concentration indicators of corner modes over time in the optimized and the classical systems.}
\end{figure*}

For the case of 2D elastic lattice, the mathematical formulation is given in Appendix S2. Similarly, we obtain the optimized modulation parameters $\mathit{d}_2^\star=\left(0.3552,0.4910,0.3537,0.4559,0.08\pi,1.44\pi\right)^\top$ and the limiting speed for adiabaticity $\Omega_{\lim}^{\star2}=7.1908\pi\times{10}^{-4}$, the modulation time $T^{\star2}=(\phi_{\rm III}-\phi_{\rm I})/\Omega_{\lim}^{\star2} = 1891$. The limiting speed for adiabaticity and modulation time for the evolution of the corner state in the classical system\cite{RN26} ($b_x=b_y=1/3,\delta_x=\delta_y=0.6$) are $\Omega_{\lim}^2=7.4070\pi\times{10}^{-5}$ and $T^2=9451$, indicating that optimized parameters can reduce the modulation time to about 1/5.0 of the original. Figs. \ref{fig:4}(a)-\ref{fig:4}(f) show the spectrums, adiabatic and non-adiabatic regions and adiabatic transition, the displacement fields, and concentration indicators of corner modes over time for the 2D optimized and the classical systems, which fully validate that the design paradigm can effectively achieve an adiabatic corner-to-corner pumping of elastic wave and significantly improve the modulation speed.

\begin{figure*}
\includegraphics[width=0.9\textwidth]{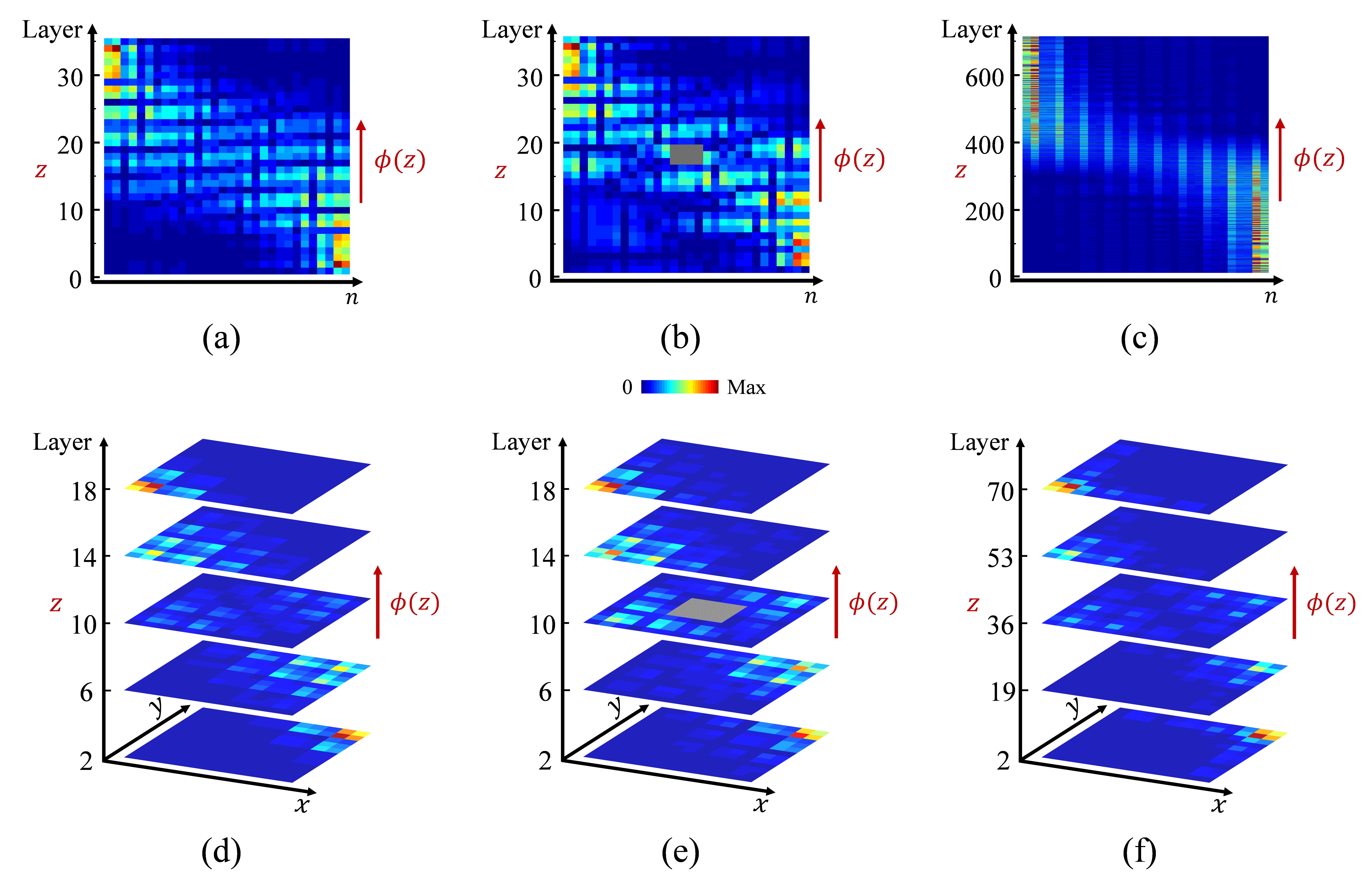}
\caption{\label{fig:5} Evaluation and robustness validation of the optimized modulation schemes applied to spatially modulated systems. (a) Edge-bulk-edge pumping in the 1D optimized systems with frequency \textit{f} = 1.7771. (b) Robustness validation of the 1D optimized system (with 3$\times$4 mass blocks removed). (c) Edge-bulk-edge pumping in the 1D classical system with \textit{f} = 2.0079. (d) Corner-bulk-corner pumping in the 2D optimized system with \textit{f} = 2.7563. (e) Robustness validation of the 2D optimized system (with 4$\times$4 mass blocks removed). (f) Corner-bulk-corner pumping in the 2D classical system with \textit{f} = 2.8954.}
\end{figure*}

Due to the strict request for modulation methods and materials of time-modulated systems in experiments, the performance of the optimized modulation schemes in the space-modulated system is also investigated, focusing on whether it can significantly reduce the number of stacks. Furthermore, spatial modulation also makes it easier to validate the immune-deficiency characteristics of the pumping systems. For a 1D spatially modulated system, the optimized system can achieve highly concentrated edge-to-edge energy transport by stacking only 35 layers when using the same parameters as the temporal modulation scheme, as shown in Fig. \ref{fig:5}(a). Even if 3 × 4 mass blocks are removed as defects in the center of the system, the optimized system still allows for smooth pumping, as shown in Fig. \ref{fig:5}(b). With the same requirement for concentration indicators of edge modes (see Appendix S3), the classical system requires 657 layers, as shown in Fig. \ref{fig:5}(c). For the 2D elastic system, pumping of corner states featured with highly concentrated energy transport can be implemented when 19 layers are stacked using the optimized modulation schemes, regardless of whether the system is perfect or with 4 × 4 mass blocks removed, as demonstrated in Figs. \ref{fig:5}(d) and \ref{fig:5}(e). Nevertheless, the classical system requires stacking 71 layers to achieve the same level of concentrated corner-to-corner energy transport, as shown in Fig. \ref{fig:5}(f). This demonstrates that the optimized modulation schemes can also be applied to space-modulated systems and can significantly reduce the number of stacks (94.7\% and 73.2\% reduction in the number of stacks for 1D and 2D pumping systems, respectively).

In this letter, a rational optimization design method is proposed for the first time for temporal pumping in 1D and 2D elastic lattices with tunable stiffness. By optimizing the modulation parameters, the pumping systems can achieve adiabatic evolution of specified edge or corner states with the fastest linear modulation speed. Through transient simulations, it is validated that the optimized modulation scheme can reduce the modulation time to about 1/12.2 and 1/5 of the original modulation time compared to the 1D and 2D classical systems while ensuring a highly concentrated energy transport. Interestingly, the space-modulated systems can also benefit from the optimized modulation schemes of the time-modulated systems. On the premise of achieving robust edge-to-edge or corner-to-corner energy transport, the optimized schemes can drastically reduce the number of stacks compared to the classical schemes, which is of great significance for experiments. The optimization design method proposed in this letter can also be extended to the rational design of fast topological pumping for other physical fields, such as acoustics and photonics.

\vbox{}

This work was financially supported by the National Natural Science Foundation of China (Nos. 11821202, 12372122 and 12302113), the Science Technology Plan of Liaoning Province (No. 2023JH2/101600044), and the Ningbo Municipal Natural Science Foundation (No. 2022J090).

\section*{AUTHOR DECLARATIONS}
\section*{Conflict of Interest} The authors have no conflicts to disclose.
\section*{Author Contributions} 
\textbf{Xianggui Ding}: Data curation (equal); Formal analysis (equal); Investigation (equal); Visualization (equal); Writing – original draft (equal). \textbf{Zongliang Du}: Conceptualization (equal); Methodology (equal); Supervision (equal); Writing – review \& editing (equal);  Funding acquisition (equal). \textbf{Jiachen Luo}: Investigation (equal); Writing – review \& editing (equal). \textbf{Hui Chen}: Methodology (equal); Supervision (equal); Writing – review \& editing (equal);  Funding acquisition (equal). \textbf{Zhenqun Guan}: Writing – review \& editing (equal). \textbf{Xu Guo}: Conceptualization (equal); Supervision (equal); Writing – review \& editing (equal);  Funding acquisition (equal). 
\section*{DATA AVAILABILITY} The data that support the findings of this study are available from the corresponding authors upon reasonable request.

\section*{REFERENCES}
\nocite{*}
\bibliography{References}
\end{document}